\documentclass[10pt]{article}
\usepackage{shadow,pifont,epsfig,wrapfig,times,graphics,color}
\usepackage{latexsym}
\usepackage{graphicx, graphics, rotating}
\usepackage{pstcol, pstricks, pst-node, pst-coil, pst-grad}

\begin{document}
\def\dsdt{$\frac{d\sigma}{dt}$}
\def\beqn{\begin{eqnarray}}
\def\eeqn{\end{eqnarray}}
\def\barr{\begin{array}}
\def\earr{\end{array}}
\def\btab{\begin{tabular}}
\def\etab{\end{tabular}}
\def\bite{\begin{itemize}}
\def\eite{\end{itemize}}
\def\bcen{\begin{center}}
\def\ecen{\end{center}}

\def\eq{\begin{equation}}
\def\ee{\end{equation}}
\def\eqa{\begin{eqnarray}}
\def\eea{\end{eqnarray}}
\def\nn{\nonumber}
\def\psmu{P^{\prime \mu}}
\def\psnu{P^{\prime \nu}}
\def\ksmu{K^{\prime \mu}}
\def\pss{P^{\prime \hspace{0.05cm}2}}
\def\psf{P^{\prime \hspace{0.05cm}4}}
\def\kdagger{K\hspace{-0.3cm}/}
\def\ndagger{N\hspace{-0.3cm}/}
\def\psdagger{p'\hspace{-0.28cm}/}
\def\epssdagger{\varepsilon'\hspace{-0.28cm}/}
\def\epsdagger{\varepsilon\hspace{-0.18cm}/}
\def\pdagger{p\hspace{-0.18cm}/}
\def\xidagger{\xi\hspace{-0.18cm}/}
\def\qsdagger{q'\hspace{-0.3cm}/}
\def\qdagger{q\hspace{-0.2cm}/}
\def\keldagger{k\hspace{-0.2cm}/}
\def\ksdagger{k'\hspace{-0.3cm}/}
\def\q2dagger{q_2\hspace{-0.35cm}/\;}
\def\qqs{q\!\cdot\!q'}
\def\lls{l\!\cdot\!l'}
\def\lp{l\!\cdot\!p}
\def\lps{l\!\cdot\!p'}
\def\lsp{l'\!\cdot\!p}
\def\lsps{l'\!\cdot\!p'}
\def\lqs{l\!\cdot\!q'}
\def\pps{p\!\cdot\!p'}
\def\psqs{p'\!\cdot\!q'}
\def\epsp{\varepsilon'\!\cdot\!p}
\def\epsps{\varepsilon'\!\cdot\!p'}
\def\epsl{\varepsilon'\!\cdot\!l}
\def\epsls{\varepsilon'\!\cdot\!l'}

\title{Electromagnetic Form Factors
and the hypercentral constituent quark model}
\author{M. De Sanctis,\\
{\small  INFN,Sezione di Roma1, Piazzale Aldo Moro, Roma (Italy) } \\
M.M. Giannini, E. Santopinto and A.Vassallo \\
{\small Universit\'a di Genova e INFN, Sezione di Genova, via Dodecaneso 33, 16142 Genova (Italy)}\\
}                     
\maketitle
\abstract{
We briefly report on results about the electromagnetic form factors of the 
nucleon obtained with different models and then we concentrate our attention 
on recent results obtained with the hypercentral constituent quark model 
(hCQM).\\
}\\ 

\section{Introduction}
\label{intro}
The new data on the ratio of the electric and magnetic form factors of the
proton \cite{Jones:1999,Gayou:2001}showing an unexpected decrease with
$Q^2$ have triggered again the
interest in the description
of the internal nucleon structure in terms of various effective 
models: bag models, chiral soliton models, quark-diquark, constituent
quark models, etc.. The proton has an excitation spectrum and a finite
size: these two properties are strictly related and are both an indication 
of the composite character of the proton. 
Already in 1973 Iachello, Jackson and Land\'e \cite{Iachello:1972nu}
 were able to obtain a good
reproduction of all the existing nucleon form factors data using a 
Vector Meson Dominance (VMD) model introducing an
intrinsic form factor to describe the internal structure of the nucleon.
The results of the original fit, if one plots the ratio of $G_E/G_M$, show
not only a decrease with $Q^2$ but also a crossing of the zero at about
$8÷GeV^2$.  In 1995 using a constituent quark model Cardarelli et al.
\cite{Cardarelli:1995dc} have
calculated 
the e.m. form factors of the nucleon in a
light front approach fitting the SLAC data \cite{Hohler,Walker} by means
of form
factors for the constituent quarks.
Frank et al. in 1996
\cite{Frank:1995pv} have
constructed a relativistic light cone constituent quark model and 
calculated the electric and magnetic form factors of the proton. 
If one plots their original results  as a ratio of
the electric and magnetic form factors one  can see a strong decrease with
$Q^2$ due to the presence of a zero in the 
electric form factor at $Q^2~=~6~GeV^2$. In 2002 Miller  
\cite{Miller:2002} with
a refined version of the model has improved the reproduction of the  
decrease with $Q^2$ of the ratio.
In 1999 \cite{ts99} with a simple non relativistic quark model, the hCQM
\cite{pl},
boosting the
initial and final state to the Breit Frame and considering relativistic
corrections to the non relativistic current \cite{mds} we have shown
explicitly that
the
decrease is a relativistic effect \cite{ts99} and it disappears without
these
corrections
\cite{ts99,fra99,rap}. This calculation makes use of the nucleon form
factors previously determined \cite{mds}. Using a chiral CQM and a point
form dynamics the Pavia-Graz
group \cite{Wagenbrunn:2000es,Boffi:2001zb} has shown a good
reproduction of the form
factors and of
the ratio up to $4 GeV^2$. In 1996 Holzwarth \cite{Holzwarth:1996xq} has shown
that  the simple Skyrme soliton model, with vector meson corrections and with
the
nucleon initial and final states  boosted to the
Breit Frame, leads to $G^p_E$ that decreases with $Q^2$ and  
crosses zero at $10~GeV^2$. 
 In the MIT Bag model it is expected a sharp decrese and a zero at
$Q^2~=1.5 GeV^2$ with a change of sign, but with a  cloudy bag model
it is shown that the inclusion of the pion cloud not only  improves the
static properties of the model and  restore the chiral symmetry but also
improves the behavior of the ratio $G^p_E/G^p_M$
\cite{Lu:1997sd,Lu:1999np,Lu:yc}. 
 
Finally we can say that the extended VMD model by Lomon
\cite{Lomon:2002jx},
the soliton
model calculation by Holzwarth \cite{Holzwarth:1996xq}, the calculation by
Miller \cite{Miller:2002}
 and the relativistic quark spectator-diquark model calculation by Ma et
al.\cite{Ma} describe the new Jlab data quite well.

In the following we will concentrate our attention on the results
obtained with a very simple CQM, the hypercentral constituent quark model 
\cite{pl}. We shall introduce the model and then we shall show the new
results about the e.m. form factors  obtained with a relativistic version
of the model and a relativistic current.

\section{The hypercentral model} 
\label{sec:1} 
\begin{figure} 
\resizebox{1.\textwidth}{!}{%
  \includegraphics{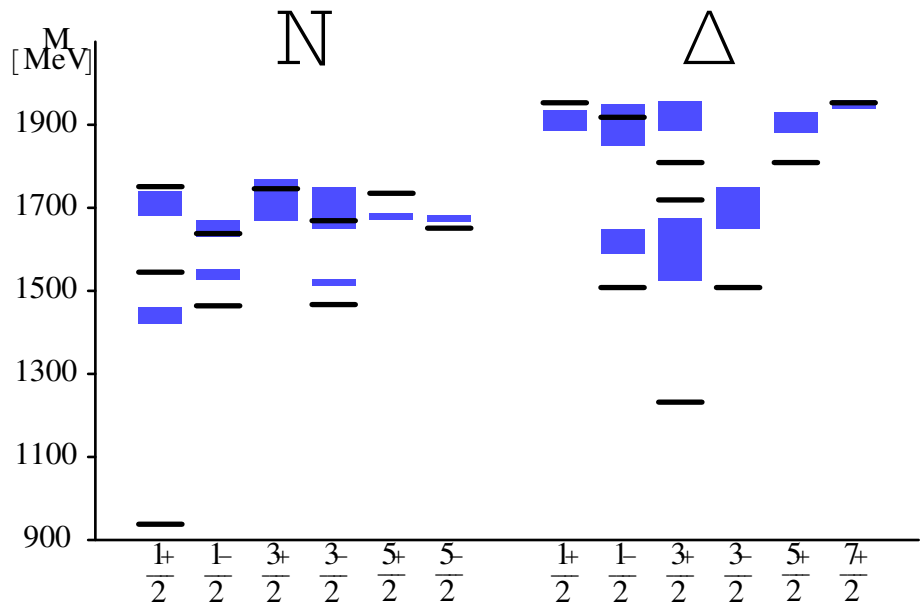}} 
\caption{The spectrum obtained with the hypercentral model Eq.(3) and the 
parameters Eq. (4) ( full lines)), compared with the experimental data of 
PDG \cite{pdg} (grey boxes).}
\label{spettro_a_t}
\end{figure}

The experimental $4$ and $3$ star non strange resonances can be arranged
in  $SU(6)$ multiplets. This means that the quark dynamics
has a dominant $SU(6)-$ invariant part, which accounts for the average
multiplet energies. In the hCQM it is assumed to be
\cite{pl}
\begin{equation}\label{eq:pot}
V(x)= -\frac{\tau}{x}~+~\alpha x,
\end{equation}
where $x$ is the hyperradius 
\begin{equation}
x=\sqrt{{\vec{\rho}}^2+{\vec{\lambda}}^2} ~~,
\end{equation}
where $\vec{\rho}$ and $\vec{\lambda}$ are the Jacobi coordinates
describing the
internal quark motion. The dependence of the potential on the hyperangle
$\xi=arctg(\frac{{\rho}}{{\lambda}})$ has been neglected.\\
Interactions of the type linear plus Coulomb-like have been used since
long time for the meson sector, e.g. the Cornell potential. This form has been
supported by recent Lattice QCD calculations \cite{bali}.\\
In the case of
baryons a so called hypercentral approximation has been
introduced \cite{has,rich}, this approximation amounts to average 
any two-body 
potential for the three quark system over the hyperangle $\xi$ and works
quite well, specially for the lower part of the spectrum \cite{hca}. In
this respect, the hypercentral potential Eq.\ref{eq:pot} can be considered
as the hypercentral approximation of the Lattice QCD potential. On the
other hand, the hyperradius $x$ is a collective coordinate and therefore
the hypercentral potential contains also three-body effects.\\
The hypercoulomb term $1/x$ has important
properties \cite{pl,sig}: it can be solved analytically and the resulting
form factors have a power-law behaviour, at variance with the widely used
harmonic oscillator; moreover, the negative parity states are exactly
degenerate with the first positive parity excitation, providing a good
starting point for the description of the spectrum.\\ 
The splittings within the multiplets are produced by a perturbative term
breaking the $SU(6)$ symmetry, 
which, as a first approximation, can be assumed to be the
standard hyperfine interaction $H_{hyp}$ \cite{is}.
The three quark hamiltonian for the hCQM is then:
\begin{equation}\label{eq:ham}
H = \frac{p_{\lambda}^2}{2m}+\frac{p_{\rho}^2}{2m}-\frac{\tau}{x}~
+~\alpha x+H_{hyp},
\end{equation}
where $m$ is the quark mass (taken equal to $1/3$ of the nucleon mass). 
The strength of the hyperfine interaction is determined in order to
reproduce the $\Delta-N$ mass difference, the remaining two free
parameters are fitted to the spectrum, reported in Fig.\ref{spettro_a_t}, 
leading to the following values $\alpha= 1.61~fm^{-2},~~~~\tau=4.59~$.

Keeping these parameters fixed, the model has been applied to calculate 
various physical quantities of interest: the photocouplings \cite{aie},
the
electromagnetic transition amplitudes \cite{aie2}, the elastic nucleon
form factors \cite{mds} and the ratio between the electric and magnetic form
factors of the proton \cite{rap}.

\section{The results}
\label{sec:2} 
 The electromagnetic transition amplitudes 
are defined as the matrix elements of the  
electromagnetic interaction, 
between the nucleon, $N$, and the resonance, $B$, states:
\begin{equation}\label{eq:amp}
\begin{array}{rcl}
A_{1/2}&=& \langle B, J', J'_{z}=\frac{1}{2}\ | H_{em}^t| N, J~=~
\frac{1}{2}, J_{z}= -\frac{1}{2}\
\rangle\zeta\\
& & \\
A_{3/2}&=& \langle B, J', J'_{z}=\frac{3}{2}\ | H_{em}^t| N, J~=~
\frac{1}{2}, J_{z}= \frac{1}{2}~\
\rangle~\zeta~\\
& & \\
S_{1/2}&=& \langle B, J', J'_{z}=\frac{1}{2}\ | H_{em}^l| N, J~=~
\frac{1}{2}, J_{z}= \frac{1}{2}~\
\rangle~\zeta~\\ 
\end{array}
\end{equation}
where $\zeta$ is the sign of the N$\pi$ amplitude.

\begin{figure}[h] 
\resizebox{1.0\textwidth}{!}{%
\includegraphics[angle=90]{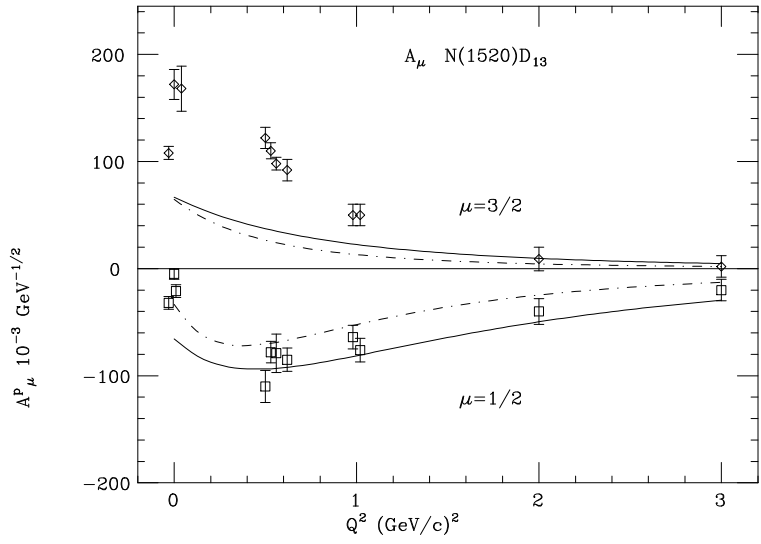} 
} 
\caption{
The helicity amplitudes for the $D_{13}(1520)$ resonance,
calculated with the hCQM of Eqs. (3) and (4) (full curve, \cite{aie2}).
The dashed curve is obtained with the analytical version of the hCQM
(\cite{sig}), where the behaviour of the quark wave function is determined
mainly by the hypercoulomb potential. The data are from the compilation of
ref. \cite{burk}
}
\label{d13}       
\end{figure}

\begin{figure}[hb] 
\resizebox{1.0\textwidth}{!}{%
\includegraphics{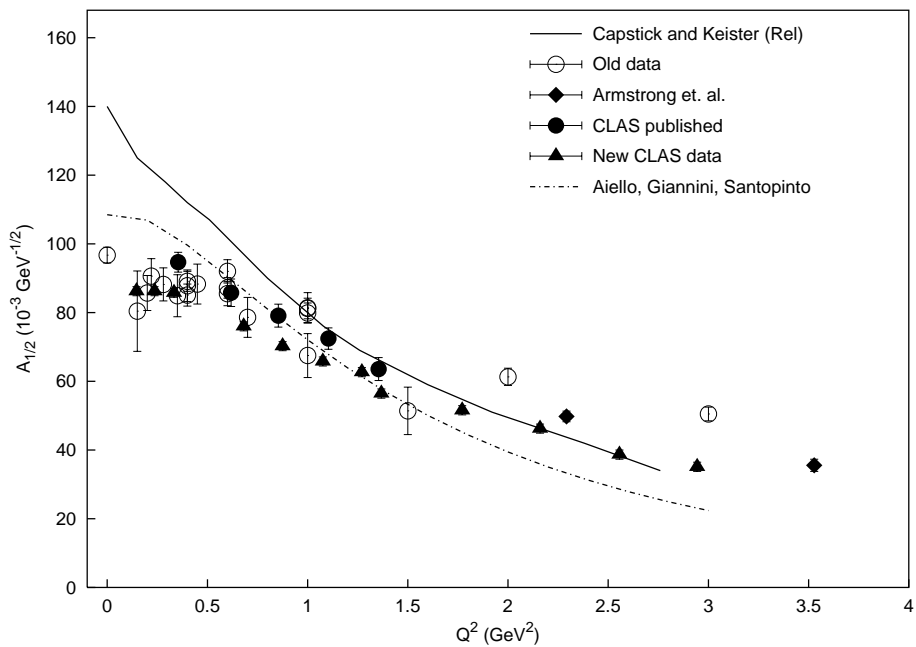} 
} 
\caption{
The helicity amplitudes for the $S_{11}(1535)$ resonance,
calculated with the hCQM of Eqs. (3) and (4) (dotted curve, \cite{aie2}) 
and the model of ref.
\cite{cap} (full curve). 
The data are taken from the compilation of ref. \cite{burk2}
}
\label{s11}       
\end{figure}

The proton photocouplings of the hCQM \cite{aie} (Eq.~(\ref{eq:amp})
calculated  at the photon point), in comparison with other
calculations \cite{ki,cap,bil}, have the same overall behaviour,
having the same SU(6)
structure in common, and  they all show a lack of strength: considering 
the square modulus of the helicity amplitudes these discrepancies, on  
average, range from 30 to 50 \%.\\
Taking into account the $Q^2$ behaviour of the transition matrix elements 
of Eq.~(\ref{eq:amp}), one can calculate the hCQM helicity amplitudes in
the Breit frame \cite{aie2}. The hCQM results for the $D_{13}(1520)$ and
the $S_{11}(1535)$ resonances \cite{aie2} are given in Fig.\ref{d13} and 
\ref{s11},
respectively. The agreement in the case of the $S_{11}$ is remarkable, the
more so since the hCQM curve has been published three years in advance
with respect to the recent TJNAF data \cite{dytman}.
We have also calculated the longitudinal helicity amplitudes \cite{long}. 
In general the $Q^2$ behaviour is reproduced, except for
~~~discrepancies at small $Q^2$, especially in the
$A^{p}_{3/2}$ amplitude of the transition to the $D_{13}(1520)$ state.
 The kinematical relativistic
corrections at the level of boosting the nucleon  and the resonance  
states to a common frame are not responsible for these discrepancies,  as
we have demonstrated in Ref.~\cite{mds2}.  
These discrepancies, as the ones observed in the photocouplings, can be
ascribed  to the lack of explicit quark-antiquark configurations, 
which may be important at low $Q^{2}$: in reference \cite{lothar} is 
shown that pion loops are important in this $Q^2$ region.  
Similar results are obtained for 
the other negative parity resonances \cite{aie2}. 
It should be mentioned that the r.m.s. radius of the proton corresponding 
to the parameters of the hypercentral potential determined in the previous
section is
$0.48$ fm, which is the same  value obtained in \cite{cko} in order to
reproduce the
$D_{13}$ photocoupling. 

For example,
for the Delta resonance the contribution of the pion cloud is 
very important  \cite{lothar}. 
For the transverse amplitudes $A_{1/2}$ and $A_{3/2}$ it is 
about $50\%$ at low $Q^2$ and for the longitudinal amplitude as well as for 
the electric amplitude the pion cloud is absolutely dominant.

\section{The isospin dependence}
In the chiral Constituent Quark Model \cite{olof,ple}, the non
confining part of the   
potential is provided by the interaction with the Goldstone bosons,
giving rise to a spin- and flavour-dependent term, which is crucial in
this approach for the description of the lower part of the spectrum.
More generally, one can expect that the quark-antiquark pair production 
can lead to an effective residual quark interaction containing an isospin
(flavour) dependent term.

\begin{figure} 
\resizebox{1.\textwidth}{!}{%
\includegraphics{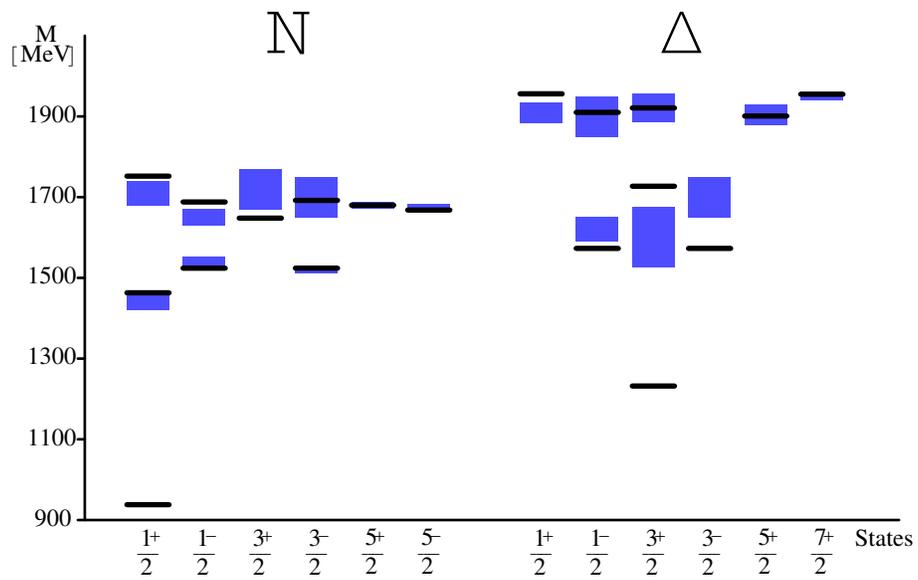} 
} 
\caption{
The spectrum obtained with the hypercentral model
containing isospin dependent terms Eq. (7) \cite{iso} (full lines)),
compared with the
experimental data of PDG \cite{pdg} (grey boxes)
}
\label{iso}       
\end{figure}

Therefore, we have introduced isospin dependent terms in the hCQM 
hamiltonian. The complete interaction used is given by \cite{iso}
\begin{equation}\label{tot}
H_{int}~=~V(x) +H_{\mathrm{S}} +H_{\mathrm{I}} +H_{\mathrm{SI}}~,
\end{equation}
where $V(x)$ is the linear plus hypercoulomb SU(6)-invariant potential of Eq.
\ref{eq:pot}, while the remaining terms are the residual SU(6)-breaking
interactions, responsible for the splittings within the multiplets. 
${H}_{\mathrm{S}}$ is a smeared standard hyperfine term,  
${H}_{\mathrm{I}}$ is isospin dependent and  ${H}_{\mathrm{SI}}$ 
spin-isospin dependent.
The resulting spectrum for the 3 and 4 star resonances is shown in Fig. 
\ref{iso} \cite{iso}. The contribution of the
hyperfine interaction to the $N-\Delta$ mass difference is in this case 
only about $35\%$, while the remaining splitting comes from the
spin-isospin term, $(50\%)$, and from the isospin one, $(15\%)$.
It should be noted that the position of the Roper and the negative 
parity states is well reproduced.

\section{Relativity}

The relativistic effects that one can introduce starting from a non
relativistic quark model are:
a) the relativistic kinetic energy;
b) the boosts from the rest frames of the initial and final baryon to a 
common (say the Breit) frame;
c) a relativistic quark current. All these features are not equivalent to 
a fully relativistic dynamics, which is still beyond the present 
capabilities of the various models.

The potential of Eq.\ref{eq:pot} has been refitted using the correct 
relativistic kinetic energy \cite{mds3}
\begin{equation}\label{eq:hrel}
H_{rel} = \sum_{i=1}^3 \sqrt{p_{i}^2+m^2}-\frac{\tau}{x}~
+~\alpha x+H_{hyp}.
\end{equation}
 The resulting spectrum is not much different from the non relativistic
one and the parameters of the potential are only slightly modified.

The boosts and a relativistic quark current expanded up to lowest order 
in the quark momenta has been used both for the elastic form factors of
the nucleon \cite{mds} and the helicity amplitudes \cite{mds2}. For
the elastic form factors, the relativistic effects are quite strong and
bring the theoretical curves much closer to the data; in any case they are
responsible for the decrease of the ratio between the electric and magnetic
proton form factors, as it has been shown in Ref.~ 
\cite{rap}, in qualitative agreement with the recent 
Jlab data \cite{ped,gay}.

\begin{figure}[hb]
\resizebox{1.0\textwidth}{!}{%
\includegraphics{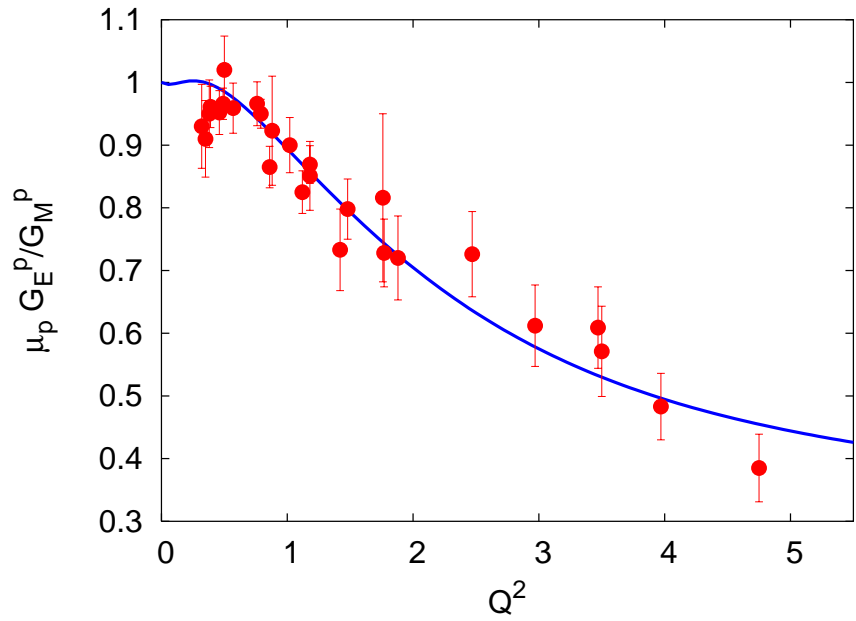}
} 
\caption{The ratio between the electric and magnetic proton form
factors, calculated with the relativistic hCQM of eq. (8),
a relativistic current and a small constituent quark form factor \cite{mds3},
 compared with the TJNAF data 
\cite{ped,gay}}
\label{rap}       
\end{figure}

A relativistic quark current, with no expansion in the quark momenta, and 
the boosts to the Breit frame have been applied to the calculation of the
elastic form factors in the relativistic version of the hCQM
Eq. (\ref{eq:hrel}) \cite{mds3}.
The resulting theoretical form factors of the proton, calculated without 
free parameters and using small quark form factors, are good and the
result for the ratio $G_E/G_M$ is reported  in Fig. \ref{rap} \cite{ff,mds3}.

\section{Conclusions}

The hCQM is a generalization to the baryon sector of the widely used
quark-antiquark
potential containing a coulomb plus a linear confining term. The three free
parameters have been adjusted to fit the spectrum \cite{pl} and then the 
model has been used for a systematic calculation of various physical
quantities: the photocouplings \cite{aie}, the helicity amplitudes for the
electromagnetic excitation of negative parity baryon resonances
\cite{aie2,mds2,long}, the elastic form factors of the nucleon
\cite{mds,mds3}
and the ratio between the electric and magnetic proton form
factors \cite{rap,ff,mds3}. The agreement with data is quite good, specially
for the helicity amplitudes, which are reproduced in the medium-high $Q^2$
behaviour, leaving some discrepancies at low (or zero) $Q^2$, where the
lacking quark-antiquark contributions are expected to be effective. 


\begin{thebibliography}{} 
%
%




\bibitem{Jones:1999}
M.~K.~Jones {\it et al.}  [Jefferson Lab Hall A Collaboration],
Phys.\ Rev.\ Lett.\  {\bf 84}, 1398 (2000)
[arXiv:nucl-ex/9910005].

\bibitem{Gayou:2001}
O.~Gayou {\it et al.}  [Jefferson Lab Hall A Collaboration],
Phys.\ Rev.\ Lett.\  {\bf 88}, 092301 (2002)
[arXiv:nucl-ex/0111010].

\bibitem{Iachello:1972nu}
F.~Iachello, A.~D.~Jackson and A.~Lande,
Phys.\ Lett.\ B {\bf 43} (1973) 191.


\bibitem{Cardarelli:1995dc}
F.~Cardarelli, E.~Pace, G.~Salme and S.~Simula,
Phys.\ Lett.\ B {\bf 357}, 267 (1995)
[arXiv:nucl-th/9507037].


\bibitem{Hohler}
G.~Hohler, E.~Pietarinen, I.~Sabba Stefanescu, F.~Borkowski, G.~G.~Simon,
V.~H.~Walther and R.~D.~Wendling,
Nucl.\ Phys.\ B {\bf 114} (1976) 505.


\bibitem{Walker}
R.~C.~Walker {\it et al.},
Phys.\ Rev.\ D {\bf 49} (1994) 5671.


\bibitem{Frank:1995pv}
M.~R.~Frank, B.~K.~Jennings and G.~A.~Miller,
Phys.\ Rev.\ C {\bf 54}, 920 (1996)
[arXiv:nucl-th/9509030].

\bibitem{Miller:2002}
G.A. Miller, Phys.\ Rev.\ C {\bf 66}, 032201(R) (2002).



\bibitem{ts99}
M.~De Sanctis, E.~Santopinto and M.~M.~Giannini,
{\it Proceedings of the  2nd ICTP International Conference on Perspectives in
Hadronic Physics, Trieste, Italy, 10-14 May 1999} World Scientific, pag. 285.


\bibitem{pl}
M. Ferraris, M.M. Giannini, M. Pizzo, E. Santopinto and L. Tiator, Phys. 
Lett. {\bf B364}, 231 (1995). 

\bibitem{mds}
M.~De Sanctis, E.~Santopinto and M.~M.~Giannini,
Eur.\ Phys.\ J.\ A {\bf 1}, 187 (1998)
[arXiv:nucl-th/9801015].

\bibitem{fra99}
M.M. Giannini, Nucl. Phys. A666 \& 667, 321c (2000).

\bibitem{rap}
M.~De~Sanctis, M.~M.~Giannini, L.~Repetto and E.~Santopinto,
Phys.\ Rev.\ C {\bf 62} (2000) 025208.



\bibitem{Wagenbrunn:2000es}
R.~F.~Wagenbrunn, S.~Boffi, W.~Klink, W.~Plessas and M.~Radici,
Phys.\ Lett.\ B {\bf 511}, 33 (2001)
[arXiv:nucl-th/0010048].

\bibitem{Boffi:2001zb}
S.~Boffi, L.~Y.~Glozman, W.~Klink, W.~Plessas, M.~Radici and
R.~F.~Wagenbrunn,
Eur.\ Phys.\ J.\ A {\bf 14}, 17 (2002)
[arXiv:hep-ph/0108271].


\bibitem{Holzwarth:1996xq}
G.~Holzwarth,
Z.\ Phys.\ A {\bf 356}, 339 (1996)
[arXiv:hep-ph/9606336].

\bibitem{Lu:1997sd}
D.~H.~Lu, A.~W.~Thomas and A.~G.~Williams,
Phys.\ Rev.\ C {\bf 57}, 2628 (1998)
[arXiv:nucl-th/9706019].

\bibitem{Lu:1999np}
D.~H.~Lu, S.~N.~Yang and A.~W.~Thomas,
J.\ Phys.\ G {\bf 26}, L75 (2000)
[arXiv:nucl-th/9911065].

\bibitem{Lu:yc}
D.~H.~Lu, S.~N.~Yang and A.~W.~Thomas,
Nucl.\ Phys.\ A {\bf 684} (2001) 296.


\bibitem{Lomon:2002jx}
E.~L.~Lomon,
Phys.\ Rev.\ C {\bf 66}, 045501 (2002)
[arXiv:nucl-th/0203081].


\bibitem{Ma}
B.Q.Ma, D.Qing and I. Schmidt, Phys. Rev.C 65, 035205, 2002; Phys. Rev. C.
66, 048201, 2002

\bibitem{bali}
G. Bali et al., Phys. Rev. {\bf D51}, 5165 (1995); G. Bali, Phys. Rept. 
{\bf 343}, 1 (2001).

\bibitem{has}
P. Hasenfratz, R.R. Horgan, J. Kuti and J.M. Richard, Phys. Lett. {\bf
B94}, 401 (1980).

\bibitem{rich}
J.-M. Richard, Phys. Rep. {\bf C 212}, 1 (1992).

\bibitem{hca}
M. Fabre de la Ripelle and J. Navarro, Ann. Phys. (N.Y.) {\bf 123}, 185 
(1979).

\bibitem{sig}
E. Santopinto, F. Iachello and M.M. Giannini, Nucl. Phys. {\bf A623}, 100c 
(1997); Eur. Phys. J. {\bf A1}, 307 (1998).

\bibitem{is}
N. Isgur and G. Karl,
Phys. Rev. {\bf D18}, 4187 (1978); {\bf D19}, 2653 (1979); {\bf D20}, 1191 
(1979); S. Godfrey and N. Isgur, Phys. Rev. {\bf D32}, 189 (1985).

\bibitem{pdg}
Particle Data Group, Eur. Phys. J. {\bf C15}, 1 (2000).

\bibitem{aie}
M. Aiello, M. Ferraris, M.M. Giannini, M. Pizzo and E. Santopinto, 
Phys. Lett. {\bf B387}, 215 (1996).

\bibitem{aie2}
M. Aiello, M. M. Giannini and E. Santopinto, J. Phys. G: Nucl. Part. Phys. 
{\bf 24}, 753 (1998).

\bibitem{burk}
V.~D.~Burkert, private communication.

\bibitem{ki}
R. Koniuk and N. Isgur, Phys. Rev. {\bf D21}, 1868 (1980).

\bibitem{cap}
S. Capstick and B.D. Keister, Phys. Rev.{\bf D 51}, 3598 (1995)

\bibitem{bil}
R. Bijker, F.~Iachello and A. Leviatan, Ann. Phys. (N.Y.) {\bf 236}, 69 (
1994).

\bibitem{dytman}
R.A. Thompson et al., Phys. Rev. Lett. 86, 1702 (2001).

\bibitem{burk2}
V.~D.~Burkert,arXiv:hep-ph/0207149.

\bibitem{long}
M. M. Giannini, E.Santopinto, A. Vassallo, to be published. 

\bibitem{mds2}
M. De Sanctis, E. Santopinto and M.M. Giannini, Eur. Phys. J. {\bf A2},
403 (1998). 

\bibitem{cko}
L. A. Copley, G. Karl and E. Obryk, Phys. Lett. {\bf 29}, 117 (1969).

\bibitem{lothar}
L.Tiator, D. Drechsel, S. Kamalov, E. Santopinto, M.M. Giannini, A. Vassallo
Eur. Phys. J. {\bf A 19}, suppl. 1, 55 (2004)

\bibitem{olof}
L. Ya. Glozman and D.O. Riska, Phys. Rep. {\bf C268}, 263 (1996).

\bibitem{ple}
L. Ya. Glozman, Z. Papp, W. Plessas, K. Varga and R. F. Wagenbrunn,
Phys. Rev. {\bf C57}, 3406 (1998); L. Ya. Glozman, W. Plessas, K. Varga
and R. F. Wagenbrunn, Phys. Rev. {\bf D58}, 094030 (1998).

\bibitem{iso}
M.M. Giannini, E. Santopinto and A, Vassallo, Eur. Phys. J. {\bf A12}, 447
(2001).

\bibitem{mds3}
M. De Sanctis, M.M. Giannini, E. Santopinto and A. Vassallo, to be
published.

\bibitem{ped}
M.K. Jones et al., Phys. Rev. Lett. {\bf B84},1398 (2000).


\bibitem{gay}
O. Gayon et al., Phys. Rev. Lett. {\bf 88}, 092301 (2002).

\bibitem{ff}
M.M.Giannini, E.Santopinto, A. Vassallo, M. De Sanctis, to be published on 
Eur.Phys.J. (2004)

\end{thebibliography}
%
 
\clearpage
\newpage

\end{document}